\documentclass[showpacs,superscriptaddress,floatfix,twocolumn,prl]{revtex4}

\usepackage{amsmath}
\usepackage[dvips]{graphicx}
\usepackage{epsfig}
\usepackage{tabularx}
\usepackage{color}

\begin{document}
\title{An ultra-low dissipation  micro-oscillator for quantum opto-mechanics}
\author{E. Serra} 
\email[Corresponding author: ]{eserra@fbk.eu}
\affiliation{Interdisciplinary Laboratory for Computational Science (LISC), FBK-University of Trento, I-38123 Povo (TN), Italy}
\affiliation{Istituto Nazionale di Fisica Nucleare (INFN), Gruppo Collegato di Trento, I-38123 Povo (TN), Italy}
\author{A. Borrielli}
\affiliation{Istituto Nazionale di Fisica Nucleare (INFN), Gruppo Collegato di Trento, I-38123 Povo (TN), Italy}
\affiliation{Institute of Materials for Electronics and Magnetism, Nanoscience-Trento-FBK Division, I-38123 Povo (TN), Italy} 
\author{F.S. Cataliotti}
\affiliation{Dipartimento di Energetica, Universit\`a di Firenze, Via Santa Marta 3, I-50139 Firenze, Italy}
\affiliation{European Laboratory for Non-Linear Spectroscopy (LENS), Via Carrara 1, I-50019 Sesto Fiorentino (FI), Italy}
\affiliation{INFN, Sezione di Firenze, Via Sansone 1, I-50019 Sesto Fiorentino (FI), Italy}
\author{F.  Marin}
\affiliation{European Laboratory for Non-Linear Spectroscopy (LENS), Via Carrara 1, I-50019 Sesto Fiorentino (FI), Italy}
\affiliation{INFN, Sezione di Firenze, Via Sansone 1, I-50019 Sesto Fiorentino (FI), Italy}
\affiliation{Dipartimento di Fisica, Universit\`a di Firenze, Via Sansone 1, I-50019 Sesto Fiorentino (FI), Italy}
\author{F.  Marino}
\affiliation{European Laboratory for Non-Linear Spectroscopy (LENS), Via Carrara 1, I-50019 Sesto Fiorentino (FI), Italy}
\affiliation{INFN, Sezione di Firenze, Via Sansone 1, I-50019 Sesto Fiorentino (FI), Italy}
\affiliation{CNR-Istituto dei Sistemi Complessi, Via Madonna del Piano 10, I-50019 Sesto Fiorentino (FI), Italy}
\author{A.  Pontin}
\author{G.A.  Prodi}
\affiliation{Istituto Nazionale di Fisica Nucleare (INFN), Gruppo Collegato di Trento, I-38123 Povo (TN), Italy}
\affiliation{Dipartimento di Fisica, Universit\`a di Trento, I-38123 Povo (TN), Italy}
\author{M. Bonaldi}
\email[Corresponding author: ]{bonaldi@science.unitn.it}
\affiliation{Istituto Nazionale di Fisica Nucleare (INFN), Gruppo Collegato di Trento, I-38123 Povo (TN), Italy}
\affiliation{Institute of Materials for Electronics and Magnetism, Nanoscience-Trento-FBK Division, I-38123 Povo (TN), Italy}

%\date{\today}

\begin{abstract}
Generating non-classical states of light by opto-mechanical coupling
depends critically on the mechanical and
optical properties of micro-oscillators and on the minimization of thermal noise. We present
an oscillating micro-mirror with a mechanical quality factor $Q = 2.6\times10^6$ at cryogenic temperature and a Finesse of
65000, obtained thanks to an innovative approach to the design
and the control of mechanical dissipation. Already at 4 K with an input laser power of 2 mW, the
radiation-pressure quantum fluctuations become the main noise source,
overcoming thermal noise. This feature makes our devices particularly suitable for 
the production of pondero-motive squeezing.
\end{abstract}


\maketitle

Radiation-pressure coupling between light and a macroscopic body is a key phenomenon right on the threshold between classical and quantum vision of reality \cite{kippenberg08,favero,marquardt}. Recent experiments have indeed demonstrated quantum effects in the behavior of macroscopic objects, thanks to the cooling of mechanical oscillators down to a thermal energy below the level of their fundamental quantum state \cite{oconnell10,teufel11,chan11}. However, all the results obtained till now can just be described using a classical view of radiation, while extremely interesting effects are expected as a consequence of the peculiar quantum properties of light. For instance, an observation of quantum properties of the electromagnetic field in its mechanical interaction with the macroscopic oscillator is still lacking, and reported experimental results are limited to preliminary classical studies \cite{verlot,marino}. A major goal of this branch of the research is the production of non-classical states of light (pondero-motive squeezing  \cite{fabre94,mancini94,corbitt,clerk}, recently observed for the first time using a cloud of cold atoms as oscillator \cite{Brooks}). Moreover, in spite of the obtained strong coupling between radiation and mechanical modes \cite{groblacher09a,verhagen12}, quantum entanglement is yet to be demonstrated.

The observation of quantum opto-mechanical effects depends critically on the mechanical and optical properties of micro-oscillators \cite{anetsberger08,cole11}. The crucial factor is in any case the minimization of thermal noise. 
A typical parameter that evaluates the importance of thermal noise in an oscillator with resonance frequency $\Omega_m/2\pi$  and quality factor $Q$, kept at temperature $T$, is $n_T=\frac{k_B\,T}{\hbar\,\Omega_m}\frac{1}{Q}$. For instance, $n_T$  quantifies the oscillator occupation number that can be reached by optical cooling. 
Here we present a novel kind of oscillating micro-mirrors, for which we have measured $n_T \simeq 1$ at 10 K. As a consequence our oscillator could enter the quantum regime by exploiting radiation-pressure cooling, starting from a background temperature of 4 K (i.e., without the necessity of ultra-cryogenic operation), a goal reached at present only by photonic crystal nano-structures \cite{chan11}. 

The main strategy to reduce thermal noise is aiming to obtain a high $Q$. Many experiments have demonstrated that mechanical resonators made of silicon crystals ($10 \times 10 \times 10$ cm$^3$) can show at low temperatures loss angles as small as $Q^{-1}=10^{-9}$. Since for smaller systems this figure reduces proportionally to their characteristic size (either thickness or width, whichever is smaller) \cite{mohanty02,liu05,zendri08}, the expected loss angle for a device with a thickness around
100 $\mu$m is $Q^{-1}=10^{-6}$. Unfortunately many effects degrade the performances of a resonator, the most common being the energy loss caused by the clamping system and by the optical coating, and thermoelastic effect \cite{Suppl}. At liquid helium temperature the thermoelastic loss becomes negligible, while the loss of the high reflectivity coating remains of the order of $\phi_c=5\times 10^{-4}$ \cite{yamamoto06}. Clamping loss is usually independent from the temperature.

Our micro-mirrors devices combine three main features specifically designed to keep under control these losses. First, the oscillators are supported from the nodal points of the resonant mode under study, to minimize  the reaction forces at the mount point and obtain a clamping loss almost independent from the mechanical impedance of the holder and its internal dissipation. 
Very good performances have been obtained in nodally suspended resonators with a characteristic length of 20 mm \cite{spiel01,borrielli11}, and more recently the advantages of nodal suspensions have been demonstrated also for micro-scale devices \cite{cole11} in the "phonon tunneling" theoretical framework. Second, the mirror is positioned on the resonator on a part moving as rigidly as possible, to reduce the strain energy stored in the coating layer that is a major source of mechanical dissipation in oscillating micro-mirrors \cite{serra12,serra12b}. 
This feature also allows for relatively large mirror diameters, giving negligible diffraction losses and thus facilitating the achievement of an high Finesse. Third, each resonator is connected to the silicon wafer through a suspended frame, acting as an isolation stage that filters out seismic and other technical noise and reduces the coupling between the resonant modes of the main oscillator and the internal modes of the wafer and the sample holder. A complete device may be thought as made of three parts (resonator, coating and frame) and its overall mechanical performances are given by the interplay between the properties of each part.

\begin{figure}[ht!]
\includegraphics[width=86mm]{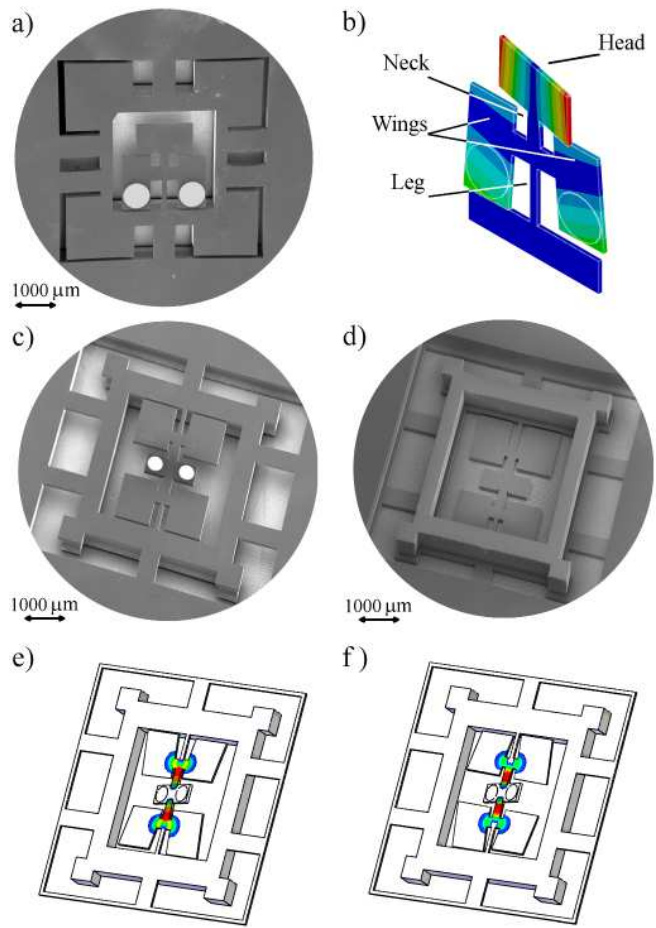}
\caption{(Color online) DPO and QPO oscillators. a) SEM image of a DPO optical resonator with a couple of mirrors on the wings. The suspended outer frame is realized from the full thickness of the wafer, and it is suspended by beams of thickness 70 $\mu$m. The four dead weights at the corners are used to tune the resonance frequencies of the suspension, whose size has been chosen to avoid the superposition of frame modes with resonator modes: the first resonant frequency of the system frame+beams is about 25 kHz, while the internal frequencies of the frame are well above 100 kHz. b) Mode-shape plot of the AS1 mode of the DPO. c) SEM image (front view) of a QPO with mirrors on the head. d) Back view. e) Mode-shape plot of the AS1 mode. f) Mode-shape plot of the AS2 mode.  The contour graph plotted over the modal shapes shows the average elastic energy stored in the device during an oscillation cycle (relative values of energy from 0 to the maximum value).\label{fig:v35v31}}
\end{figure}

Here we present two different kinds of resonators. The first one 
is based on the Double Paddle Oscillator (DPO) design \cite{spiel01}, and consists of two inertial
members, head and a couple of wings, that are connected by a torsion rod, called neck (Fig. 1a-b). The wings
are connected to the outer frame by another torsion rod, the leg. This system can
be visualized as a coupled oscillator consisting of two masses (head and wings) and of
two springs (neck and leg) that twist or bend in different directions, originating several
composite vibration modes \cite{borrielli11}. The Antisymmetric torsion modes (AS) consist of a twist of the neck around the DPO symmetry axis and a synchronous oscillation of the wings. 
The oscillations of head and wings can be in phase (AS1 mode) or out of phase (AS2 mode). For these AS modes the elastic energy is primarily located at the neck, where the maximum strain field occurs during the oscillations, while the leg remains at rest and the foot can be supported by the outer frame  with negligible energy dissipation. 

Our second resonator, the Quad Paddle Oscillator (QPO), is an innovative design  derived form the DPO. It consists of three inertial members, head and two couples of wings, connected by the neck torsion rod (Fig. 1c-f).  Here again, some modes induce only a very small strain in the legs and can be supported by the outer frame with a negligible energy dissipation. We are dealing in particular with the anti-symmetric modes, where the oscillations of head and wings can be in phase (AS1 mode) or out of phase (AS2 mode). Due of the considerable modal density of this device, special care has been taken to avoid the superposition of the AS with a low-Q flexural mode. 
We note that, thanks to the relatively thick structures and the high thermal conductivity of silicon at cryogenic temperature, this device can manage input power of about 10 mW at 4 K (corresponding to several hundreds of W inside the cavity) with a temperature spread less than 0.1 K  within the  oscillating  parts \cite{Suppl}. At lower temperatures, an input power of 1 mW sets a limit working temperature  of about 2 K.

\begingroup
\squeezetable
\begin{table*}
\caption{Parameters of the micro-oscillators characterized in this work. The parameters are measured at room and cryogenic temperature, and compared with the design specifications calculated with finite elements (FEM) simulations. The values of $Q$ given by FEM refer to cryogenic temperature and are calculated assuming a loss angle of $\phi_c = 5 \times 10^{-4}$ in the coating. \label{table1}}
\begin{ruledtabular}
\begin{tabular}{ll@{\hspace{5mm}}ccc@{\hspace{5mm}}cccc@{\hspace{5mm}}ccc}
% & & & FEM& & &  Room T && & & Low T \\
\multicolumn{2}{c}{ }&\multicolumn{3}{c}{FEM}&\multicolumn{4}{c}{Room T}&\multicolumn{3}{c}{Low T}\\ 
\cline{3-5}
\cline{6-9}
\cline{10-12}
 & &   Freq.  &Mass       &$Q$        & Freq.     & $A$       &Mass       &    $Q$    & $A$& $T$& $Q$        \\
 & &   (kHz)   &(kg)       &           & (kHz)      &(m$^2)$    &(kg)       &           &(m$^2$) & (K) &           \\
 & &$\pm$ 5\% &$\pm$ 10\% &$\pm$ 10\%	&	$\pm$ 1\% &$\pm$ 30\% &$\pm$ 10\% &$\pm$ 10\% & $\pm$ 30\% &  $\pm$ 30\%  &$\pm$ 10\% \\
\hline
QPO & AS1 & 67.0 & 2.2$\times 10^{-7}$ & 4.0$\times 10^{6}$ &65.1 &9.3$\times 10^{-26}$& 2.6$\times 10^{-7}$ & 0.87$\times 10^{5}$ &2.5$\times 10^{-27}$& { }8& 0.85$\times 10^{6}$ \\
& AS2 & 89.0 & 2.2$\times 10^{-7}$ & 2.0$\times 10^{6}$ & 85.5 &4.8$\times 10^{-26}$& 3.0$\times 10^{-7}$ & 1.45$\times 10^{5}$ &2.2$\times 10^{-27}$& 14 & 2.60$\times 10^{6}$ \\
\hline
DPO & AS1 & 30.0 & 1.0$\times 10^{-6}$ & 2.5$\times 10^{6}$ & 30.4 &1.0$\times 10^{-25}$& 1.1$\times 10^{-6}$ & 0.68$\times 10^{5}$ &3.3$\times 10^{-27}$&  10 & 1.10$\times 10^{6}$\\
& AS2 & 47.0 & 1.3$\times 10^{-6}$ & 1.2$\times 10^{6}$ & 46.2 &3.2$\times 10^{-26}$& 1.6$\times 10^{-6}$&  0.87$\times 10^{5}$&1.2$\times 10^{-27}$& 11 &  0.85$\times 10^{6}$ \\
\hline
DPO & AS1 & 75.0 & 1.1$\times 10^{-7}$ & 1.2$\times 10^{6}$ & 70.9 &2.3$\times 10^{-25}$& 0.9$\times 10^{-7}$ & 1.45$\times 10^{5}$ \\
& AS2 & 88.0 & 6.7$\times 10^{-7}$ & 4.0$\times 10^{6}$ & 86.4 &1.3$\times 10^{-26}$& 1.1$\times 10^{-6}$ & 1.50$\times 10^{5}$ \\ 
\end{tabular}
\end{ruledtabular}
\end{table*}
\endgroup

\begin{figure}[b!]
\includegraphics[width=86mm]{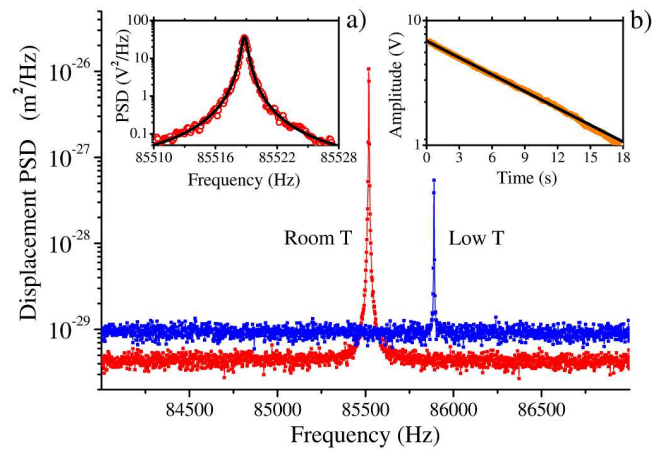}
\caption{(Color online) Main panel: power spectral density around the peak corresponding to the AS2 mode of the QPO, at room (red trace) and cryogenic (blue trace) temperature. a) The spectrum is frequency-shifted by a digital lock-in amplifier. The solid line shows the fitting function composed of a mechanical resonance plus a flat background. b) The amplitude of an excited oscillation is shown (symbols), together with the fitting exponential decay (straight line). \label{fig2}}
\end{figure}

We have used a Michelson interferometer with a Nd:YAG laser source to characterize at room temperature and at 10~K two oscillators, one DPO with 800 $\mu$m diameter mirrors on the wings and one QPO with 400 $\mu$m diameter mirrors on the head, built on the same $35 \times 35$ mm$^2$ wafer sector. We have also characterized at room temperature one further device, a different DPO with mirrors on the head. 
A portion of displacement noise spectra for the QPO is shown in Fig. \ref{fig2}. The areas below the peaks of the AS modes are measured with an accuracy of about 30$\%$, both at room and at cryogenic temperature \cite{Suppl}. We obtain a good agreement between the deduced values of the effective mass and the FEM simulations, and we also confirm that the modal temperature in the cryogenic environment corresponds to the cold finger temperature of 10~K. The results are summarized in Table I.   

The measurement of the mechanical quality factor $Q$ at room temperature is performed from the width of the spectral resonances. At this  purpose, the spectrum is frequency-shifted by a digital lock-in amplifier, whose internal local oscillator is tuned at 110 Hz from the peak. The beat note is then recorded by a digital scope and analyzed by FFT (Fig. \ref{fig2}a). At cryogenic temperature we have adopted a time-resolved detection technique. An excitation beam is amplitude modulated at a frequency very close to the mechanical resonance, then the modulation is switched off and the amplitude of the mechanical oscillation (monitored by a Michelson interferometer) is measured by the lock-in amplifier whose output quadratures are acquired by the scope. The oscillation amplitude is then calculated off-line and $Q$ is evaluated from the decay time (Fig. \ref{fig2}b). The results are summarized in Table I and show that the very high values planned by the design could indeed be achieved.

After the  measurements with the Michelson interferometer, we have formed a 0.6~mm long Fabry-Perot cavity using one of the 400 $\mu$m mirrors on the head of the QPO as end mirror, and a $50$~mm radius silica input mirror (its intensity transmittance is $5\times 10^{-5}$). The input coupler is fixed on a piezo-electric transducer for cavity tuning.  The cavity has a waist of $43\, \mu$m, allowing negligible diffraction losses even on the micro-mirror, and we have measured \cite{Suppl} a Finesse of 65000, at room temperature, that is the best value ever reported for oscillating micro-mirrors.

In order to assess an oscillating micro-mirror with mass $M$ in view of the production of pondero-motive squeezing, the main criterion is the comparison between the power spectral density (PSD) of the thermal noise force  $\,\,S_T = 2\,k_B\,T\frac{M \Omega_m}{Q}\,\,$ with the PSD of the radiation pressure due to quantum fluctuations. The shot-noise of the input field is amplified in the cavity and gives rise to a force spectral density expressed as $\,\,S_{RP} = \hbar \omega_L P_{in}\frac{4}{c^2} \left(\frac{2\, \mathcal{T}}{\mathcal{T}+\mathcal{L}}\right)^2 \left(\frac{\mathcal{F}}{\pi}\right)^2 \,\,$ ($P_{in}$ is the input laser power, $\omega_L/2\pi$ is the laser frequency, $\mathcal{F} = 2 \pi/(\mathcal{T}+\mathcal{L})$ is the cavity Finesse, $\mathcal{T}$ and $\mathcal{L}$ are respectively the input mirror intensity transmission and the cavity losses). This expression is valid for vanishing detuning and in the bad-cavity limit (cavity linewidth much larger than the oscillation frequency). In our case we have a record high $Q$, and a relatively large mass that is balanced  
by two useful features: the high Finesse and the possibility to manage high power at low temperature. Therefore, already at 4.5 K with an input laser power of 2 mW, we may obtain $\,\,S_{RP} > S_T$: the radiation-pressure force noise due to quantum fluctuations dominates over thermal noise, an effect that has not yet been observed. In this regime the generation of squeezed light can be obtained as a result of the quantum opto-mechanical correlations between field quadratures.

\begin{figure}[t!]
\includegraphics[width=86mm]{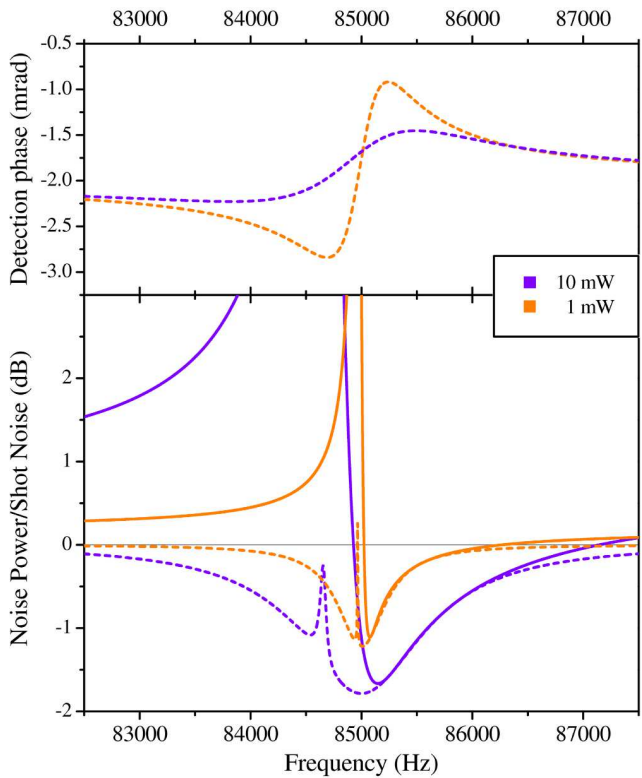}
\caption{(Color online) Lower panel: solid lines show noise spectra, normalized to the shot-noise level, calculated using the parameters given in the 
text and $P_{in}=1$~mW (orange) and $P_{in}=10$~mW (violet). The corresponding optimum squeezing 
curves (dashed lines) indicate the maximum quantum noise reduction achievable as the detection phase is optimized at each frequency.
Upper panel: detection phase necessary to obtain the optimum squeezing curves.
\label{sqteo}}
\end{figure}

For a more quantitative calculation, we have used the model described in Ref. \cite{fabre94}, extended to include cavity losses, laser extra-noise and background displacement noise. Two examples of the theoretical spectra corresponding to $P_{in}=1$~mW and $P_{in}=10$~mW are shown in Fig. (\ref{sqteo}).  We remark that here we  use  a completely measured set of experimental parameters, differently to simulations previously reported in the literature \cite{fabre94,mancini94,marino}. 
The cavity and oscillator parameters measured in this work for the QPO oscillator are:   $\mathcal{T}=5\times10^{-5}$, $\mathcal{L}=4.7\times10^{-5}$, cavity length 0.6 mm, $M=3\times 10^{-7}$~kg, $\Omega_m/2\pi=85$~kHz, $Q=2.6\times 10^{6}$. The oscillator temperature is fixed at 4.5 K. We have considered an input field detuning (normalized to the cavity half-linewidth) of $\,-10^{-3}$ in order to safely operate on the 'cooling' side of the resonance. Actually, even if in principle at null detuning the system is stable, an extremely small excursus on the 'heating' side would trigger a self-oscillation \cite{Arcizet06}. For the laser frequency noise we have taken a conservative value of 1 Hz$^2$/Hz; the amplitude noise is considered as ($1+P_{in}/10$mW) times larger than the shot noise. Finally, we have assumed a background displacement noise of $10^{-34}$~m$^2$/Hz, a value typically measured at room temperature with silicon micromirrors \cite{Arcizet08}, but that could reasonably decrease at cryogenic temperature.   

The observed quadrature is determined by the the detection phase, defined with respect to the input field. For the spectra shown with solid lines in the figure, this phase is chosen at $1.2$~mrad when $P_{in}=1$~mW and $1.5$~mrad when $P_{in}=10$~mW.
An observable noise reduction of about 2~dB on a bandwidth of $\sim 2$~kHz is 
expected with $P_{in}=10$~mW. For an input power decreased down to 1~mW the noise reduction is still greater than 1~dB, although in a narrower bandwidth.

Our experimental set-up is tailored for squeezing measurements in the bad-cavity regime (the half-linewidth is 1.95 MHz), but 
also the complementary regime of resolved sidebands can easily be achieved. 
For instance, a 49.4 mm long cavity would keep the same waist while lowering the cavity half-linewidth down to $\kappa/2\pi\sim23$ kHz. The resolved-sideband regime is necessary for optical cooling down to the mechanical ground state, and it is the starting point to create a cavity quantum opto-mechanical system where studying the strong coupling between macroscopic optical and mechanical variables. To do that, the coupling strength must exceed the mechanical and optical decoherence rates. The effective coupling strength is defined as $g = 2\frac{\omega_L}{L_{\textnormal{cav}}}\,x_{\textnormal{ZPF}}\,\sqrt{\overline{n}}$ ($L_{\textnormal{cav}}$ is the cavity length, $\overline{n}$ is the intra-cavity  photon number) where $x_{\textnormal{ZPF}}$ is the size of the ground state wavepacket $x_{\textnormal{ZPF}} = \sqrt{\frac{\hbar}{2 M \Omega_m}}$. In our case, $g$ can be increased thanks to the high bearable power, achieving $g/2\pi \simeq 30$ kHz for $P_{in} = 1$ mW. Moreover, the mechanical decoherence rate $\gamma_m = \Omega_m \, n_T$ is very weak ($\gamma_m/2\pi = 32$ kHz at 4 K), thanks to the extremely high $Q$. As a consequence, it is possible to obtain quantum strong-coupling (with both $g \gg \gamma_m$ and $g \gg \kappa$) in a cryogenic experiment with few mW of input laser power. 
Such a  regime has never been reached before, and would allow a variety of experiments exploring deeply quantum phenomena, including entanglement between macroscopic objects and light \cite{mancini02,pinard,vitali07}.

We acknowledge the support from the MEMS2 joint project of the Istituto Nazionale di Fisica Nucleare and Fondazione Bruno Kessler. A.B. acknowledges support from the European Research Council under the European Community's Seventh Framework Programme (FP7/2007-2013)/ERC grant agreement no. 202680.


\begin{thebibliography}{99}
\bibitem{kippenberg08}T. J. Kippenberg and K. J. Vahala, Science {\bf 321}, 1172 (2008).
\bibitem{favero}I. Favero and K. Karrai,  Nature Photonics, {\bf 3}, 201 (2009). \bibitem{marquardt}F. Marquardt and S. M. Girvin, Physics {\bf 2}, 40 (2009).
\bibitem{oconnell10}A. D. O' Connell, M. Hofheinz, M. Ansmann, R. C. Bialczak, M. Lenander, E. Lucero, M. Neeley, D. Sank, H. Wang, M. Weides, J. Wenner, J. M. Martinis, and A. N. Cleland, Nature {\bf 464}, 697 (2010). 
\bibitem{teufel11}J.D. Teufel, T. Donner, D. Li, J. W. Harlow, M. S. Allman, K. Cicak, A. J. Sirois, J.D. Whittaker, K. W. Lehnert, and R. W. Simmonds, Nature {\bf 475}, 359 (2011). 
\bibitem{chan11}J. Chan, T. P. Mayer Alegre, A. H. Safavi-Naeini, J. T. Hill, A. Krause, S. Gr$\ddot{\textrm{o}}$blacher, M. Aspelmeyer, and O. Painter, Nature {\bf 478}, 89 (2011). 
\bibitem{verlot}P. Verlot, A. Tavernarakis, T. Briant, P.-F. Cohadon, and A. Heidmann, Phys. Rev. Lett.  {\bf 102}, 103601 (2009). 
\bibitem{marino} F. Marino, F.S. Cataliotti, A. Farsi, M.S. de Cumis,  and F. Marin, Phys. Rev. Lett., {\bf 104}, 073601 (2010). 
\bibitem{fabre94} C. Fabre, M. Pinard, S. Bourzeix, A. Heidmann, E. Giacobino, and S. Reynaud, Phys. Rev. A {\bf 49}, 1337 (1994).
\bibitem{mancini94} S. Mancini and P. Tombesi, Phys. Rev. A {\bf 49}, 4055 (1994). 
\bibitem{corbitt} T. Corbitt, Y. Chen, F. Khalili, D. Ottaway, S. Vyatchanin, S. Whitcomb, and N. Mavalvala, Phys. Rev. A {\bf 73}, 023801 (2006). 
\bibitem{clerk}A. A. Clerk, F. Marquardt, and K. Jacobs, New J. Phys. {\bf 10}, 095010 (2008). 
\bibitem{Brooks}  D.W.C. Brooks,	T. Botter,	S. Schreppler, T.P. Purdy,	N.Brahms, and D.M. Stamper-Kurn, Nature, http://dx.doi.org/10.1038/nature11325  (2012). 
\bibitem{groblacher09a} S. Gr$\ddot{\textrm{o}}$blacher, K. Hammerer, M. R. Vanner, M. Aspelmeyer, Nature {\bf 460}, 724 (2009). 
\bibitem{verhagen12} E. Verhagen, S. Del\'eglise, S. Weis, A. Schliesser, T. J. Kippenberg, Nature {\bf 482}, 63 (2012). 
\bibitem{anetsberger08} G. Anetsberger, R. Rivi\'ere, A. Schliesser,  O.  Arcizet, and T. J.  Kippenberg, 
{Nature Photon.} {\bf 2}, 627-633 (2008).
\bibitem{cole11} G. D. Cole, I. Wilson-Rae,  K. Werbach,  M.R. Vanner,  and M. Aspelmeyer,  Nature Communications, 2:231 (2011). 
\bibitem{mohanty02}P. Mohanty,  D.A. Harrington, K.L. Ekinci, Y.T. Yang, M.J. Murphy, and M.L. Roukes, Phys. Rev. B \textbf{66}, 085416 (2002). 
\bibitem{liu05}Xiao Liu, J. F. Vignola, H. J. Simpson, B. R. Lemon, B. H. Houston, and D. M. Photiadis, J. Appl. Phys. \textbf{97}, 023524 (2005).
\bibitem{zendri08} J.P. Zendri, M. Bignotto, M. Bonaldi, M. Cerdonio, L. Conti, L. Ferrario, N. Liguori, A. Maraner, E. Serra, and L. Taffarello, Rev. Sci. Inst., {\bf 79}, 033901 (2008).  
\bibitem{Suppl} More details on mechanical dissipation mechanisms and control, on device design, assessment and fabrication, on device mechanical and optical characterization, on simulated thermal properties, are reported in the Supplemental Material.
\bibitem{yamamoto06}K. Yamamoto, S. Miyoki, T. Uchiyama, H. Ishitsuka, M. Ohashi, K. Kuroda,
T. Tomaru, N. Sato, T. Suzuki,T. Haruyama, A. Yamamoto, T. Shintomi, K. Numata, K. Waseda, K. Ito, and K. Watanabe, Phys. Rev. D, {\bf 74}, 022002 (2006).  
\bibitem{spiel01} C.L. Spiel, R.O. Pohl, and A.T. Zehnder,  Rev. Sci. Inst. \textbf{72}  1482 (2001).
\bibitem{borrielli11} A. Borrielli,  M. Bonaldi,  E. Serra, A. Bagolini, and L. Conti, J. Micromech. Microeng., {\bf 21}, 065019 (2011).  
\bibitem{serra12} E. Serra, F. S. Cataliotti, F. Marin, F. Marino, A. Pontin, G.A. Prodi, and  M. Bonaldi, J. Appl. Phys. \textbf{111}, 113109 (2012).  
\bibitem{serra12b} E. Serra, A.Borrielli, F. S. Cataliotti, F. Marin, F. Marino, A. Pontin, G.A. Prodi, and  M. Bonaldi,  Appl. Phys. Lett. \textbf{101}, 071101 (2012).
\bibitem{Arcizet06} O. Arcizet, P.F. Cohadon, T. Briant, M. Pinard, and A. Heidmann, Nature {\bf 444}, 71 (2006). 
\bibitem{Arcizet08}O. Arcizet,  C. Molinelli,  T. Briant, P.F. Cohadon, A. Heidmann, J.M. Mackowski, C. Michel, L. Pinard, O. Fran\c{c}ais,  and L. Lionel Rousseau, New J. Phys., {\bf10} 125021 (2008). 
\bibitem{mancini02}S. Mancini, V. Giovannetti, D. Vitali, and P. Tombesi, Phys. Rev. Lett. {\bf 88}, 120401 (2002). 
\bibitem{pinard} M. Pinard,   A. Dantan, D. Vitali, O. Arcizet, T. Briant, and A. Heidmann,
 Europhys. Lett.  {\bf 72}, 747  (2005).
\bibitem{vitali07} D. Vitali, S. Gigan, A. Ferreira, H. R. B$\ddot{\textrm{o}}$hm, P. Tombesi, A. Guerreiro, V. Vedral, A. Zeilinger,  and M. Aspelmeyer, 
Phys. Rev. Lett., {\bf 98}, 030405 (2007).
\end{thebibliography}
\end{document}

\title{Supplemental material to:\\"An ultra-low dissipation micro-oscillator for quantum opto-mechanics"}

\maketitle
\noindent{Enrico Serra$^{1,2,\star}$,  Antonio Borrielli$^{2,3}$, Francesco S. Cataliotti$^{4,5,6}$, Francesco Marin$^{5,6,7}$, Francesco Marino$^{5,8}$, Antonio Pontin$^{2,9}$, Giovanni A. Prodi$^{2,9}$, Michele Bonaldi$^{2,3,\star}$}

\begin{quotation}
\noindent{\scriptsize{$^1$LISC, FBK-University of Trento, I-38123 Povo (Trento), Italy\\$^2$INFN, Gruppo Collegato di Trento, I-38123 Povo, Trento, Italy
\\$^3$IMEM, Nanoscience-Trento-FBK Division, I-38123 Trento, Italy
\\$^4$Dipartimento di Energetica, Universit\`a di Firenze, 
Via Santa Marta 3, I-50139 Firenze, Italy
\\$^5$LENS, Via Carrara 1, I-50019 Sesto Fiorentino (FI), Italy
\\$^6$INFN, Sezione di Firenze
\\$^7$Dipartimento di Fisica, Universit\`a di Firenze, Via Sansone 1, I-50019 Sesto Fiorentino (FI), Italy
\\$^8$CNR-ISC, Via Madonna del Piano 10, I-50019 Sesto Fiorentino (FI), Italy  
\\$^9$Dipartimento di Fisica, Universit\`a di Trento, I-38123 Povo, Trento, Italy
\\
$\star$ e-mail:  {eserra@fbk.eu} ;  {bonaldi@science.unitn.it}}}
\end{quotation}

\vspace{1cm}
{\large
This PDF file includes the following material:
\begin{itemize}
\item Section I. Design and assessment of the devices

\item  Section II. Temperature distribution in cryogenic samples

\item Section III. Fabrication of the devices

\item Section IV. Experimental apparatus

\item  Section V. Measurement techniques

\end{itemize}}

\clearpage

%\section{Design and assessment of the devices}
\begin{center}
\textbf{\large{Section I. Design and assessment of the devices}}
\end{center}
For each resonant mode, the quality factor is calculated as $Q=\phi_T^{-1}$, with $\phi_T$ the total loss angle is defined as:
\begin{equation}
\label{eq:loss}
\phi_T=  \frac{\Delta W_{T}}{2 \pi W_T}  
\end{equation}
Here  $W_T$ is the energy stored in the resonant mode and $\Delta W_{T}$ the total energy loss per oscillation cycle.
Usually  $\Delta W_{T}$ is given by a complicate volume integral over the resonator's body, as  both the energy density and the loss factor depend on the position:
\begin{equation}
\label{eq:separatedloss}
\Delta W_{T}= \int w(\mathbf{r})\phi(\mathbf{r}){d}V  
\end{equation}
where $w(\mathbf{r})$ and $\phi(\mathbf{r})$ are respectively the local strain energy density and the local loss angle.
In our case the device is made of subsystems with homogeneous loss angles, the resonator ($R$), the coating ($C$) and the frame ($F$), therefore the total energy dissipated can be separated into three contributions: 
\begin{equation}
\label{eq:separatedloss2}
\Delta W_{T}= \phi_F\int_{F} w (\mathbf{r}) {d}V +\phi_R \int_{R} w (\mathbf{r}){d}V
+\phi_C\int_{C} w (\mathbf{r}) {d}V  
\end{equation}
where the volume integrals are evaluated over each subsystem, and $\phi_F,\,\phi_R,\,\phi_C$ are respectively the loss factor of the frame, the resonator and the coating. Therefore the total loss is given by sum of the loss angles of each subsystem, weighted by the ratio of the strain energy in the subsystem to the total strain energy of the mode:
\begin{equation}
\label{eq:separatedloss3}
\phi_T = \phi_F \frac{W_F}{W_T} +  \phi_R \frac{W_R}{ W_T} +\phi_C \frac{W_C}{W_T}  \qquad .
\end{equation}

\begin{figure}[b!]
\includegraphics[width=148mm,height=42mm]{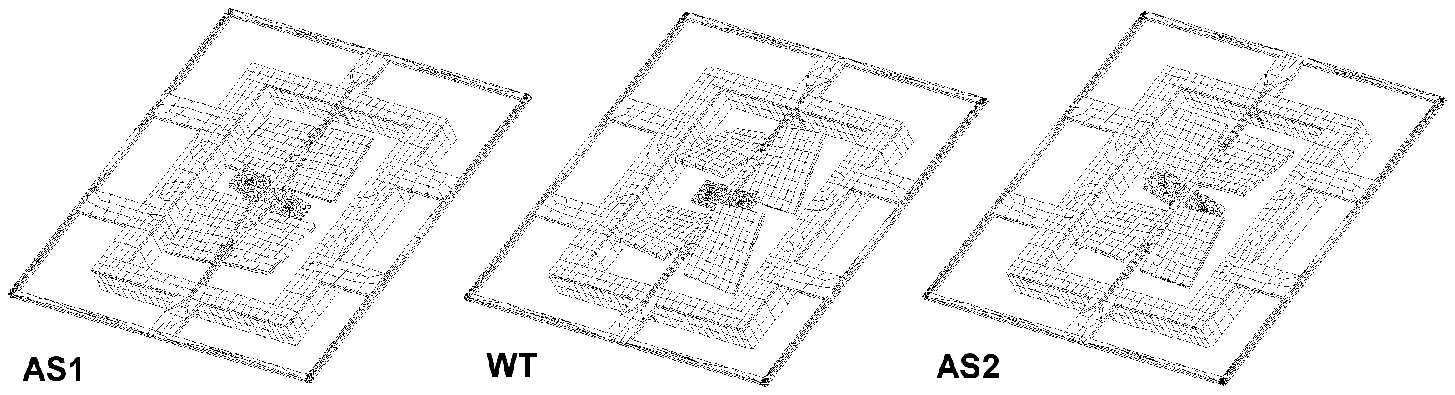}
\caption{\sffamily{Modal shapes of the QPO device.} The structure is constrained on the thin outer frame.
 \label{fig:QPOmodes}}
\end{figure}

In the design phase the geometry of each device have been optimized to reduce as much as possible the strain energy in the frame and in the coating layer for some properly selected modes of the system. We also paid attention to avoid the frequency superposition of the selected modes with low-Q flexural mode. 

Consider for instance the Quad Paddle Oscillator (QPO), consisting of three inertial members, head and two couples of wings, connected by the neck torsion rod. Here some modes induce only a very small strain in the legs and can be supported by the outer frame (F) with a negligible energy dissipation (see Figure (\ref{fig:QPOmodes})). These interesting modes are the two anti-symmetrical modes, where the oscillations of head and wings can be in phase (AS1 mode) or out of phase (AS2 mode). It is also possible to exploit  the wing torsion (WT) mode, where the head remains at rest because the neck is twisted in opposite directions by the synchronized flapping of the wings. Obviously the vibration of the WT mode can be detected only by using a mirror placed over one of the wings.

\begin{figure}[t!]
\includegraphics[width=10cm]{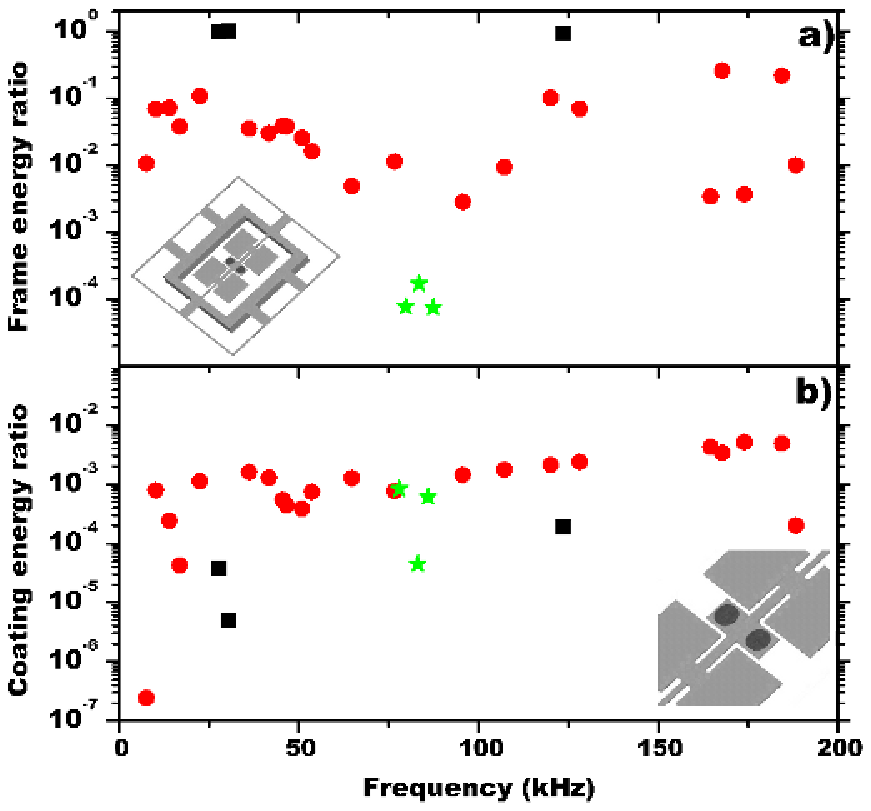}
\caption{\sffamily Strain energy distribution in a QPO device for the resonant modes below 200 kHz, as obtained by FEM. The frequencies of the AS1-WT-AS2 modes are respectively about  78-82-86 kHz. In  "frame" modes ($\blacksquare$) more than a half of the strain energy is stored in the frame, while in "resonator" modes ({\color[rgb]{1,0,0} \large{$\bullet$}}) more than a half of the strain energy is stored in the resonator. Among the resonator modes, the AS1-WT-AS2 modes are displayed as {\color[rgb]{0,1,0} \large{$\star$}}. a) Fraction of strain energy stored in the frame. Note that "frame" modes are well apart the AS and WT modes. b) Fraction of strain energy stored in the coating layer. As highlighted in the inset, two mirrors of 400 $\mu$m of diameter are deposited on both sides of the head. Since the head moves almost as a rigid body, only a maximum of about $10^{-3}$ of the oscillation energy is stored in the coating layer.
 \label{fig:v29_energy}}
\end{figure}

In the Figure (\ref{fig:v29_energy}a)  we show the  fraction of strain energy stored in the frame  for the first 30 resonant modes of a typical system resonator+frame. For most of the modes this number  ranges from 0.01 to 0.99, depending on whether the mode involves primarily the resonator or the frame. Remarkably, for the nodally suspended AS and WT modes this figure is about $10^{-4}$, i.e., orders of magnitude lower. This number means that, when we consider the  whole system resonator+frame, only a small fraction of the oscillation energy is stored in the frame and is therefore liable to be transferred to the sample holder. 
As a rule of thumb, from this distribution of strain energy we see that a mode with loss $Q^{-1}=10^{-6}$ can tolerate a loss factor of $10^{-2}$ in the frame.
In the Figure (\ref{fig:v29_energy}b) we show the fraction of strain energy stored in the coating layer for all modes.  
In this case a fraction of about $10^{-3}$ of the oscillation energy is stored in the coating, so that a mode with loss $Q^{-1}=10^{-6}$ can tolerate a loss factor of $10^{-3}$ in the coating, that is within the range of the values measured for this kind of coating   \cite{yamamoto,crooks}. 

The same device was produced also with the mirrors on the wings. In this case %, as shown in  Figure \ref{fig:v28_energy}a 
the distribution of energy between device and frame is not modified, but the tolerance to the loss factor of the coating is reduced to $10^{-4}$. This is due both to the larger size of the mirrors (800 $\mu$m) and to the larger  strain induced by the modes on the surface of the wings.

For all devices, the thermoelastic loss at room temperature of the AS and WT modes, evaluated by finite elements (FEM) simulations \cite{serra}, ranges from $\phi_{te}= 5\times 10^{-6}$ to $\phi_{te}= 1\times 10^{-5}$. We note that this loss is much lower than the thermoelastic loss of a cantilever of a similar size, thanks to the torsional character of the modes in our devices.

\clearpage
 
\begin{center}
\textbf{\large{Section II. Temperature distribution in cryogenic samples}}
\end{center}

\begin{figure}[b!]
\includegraphics[width=124mm]{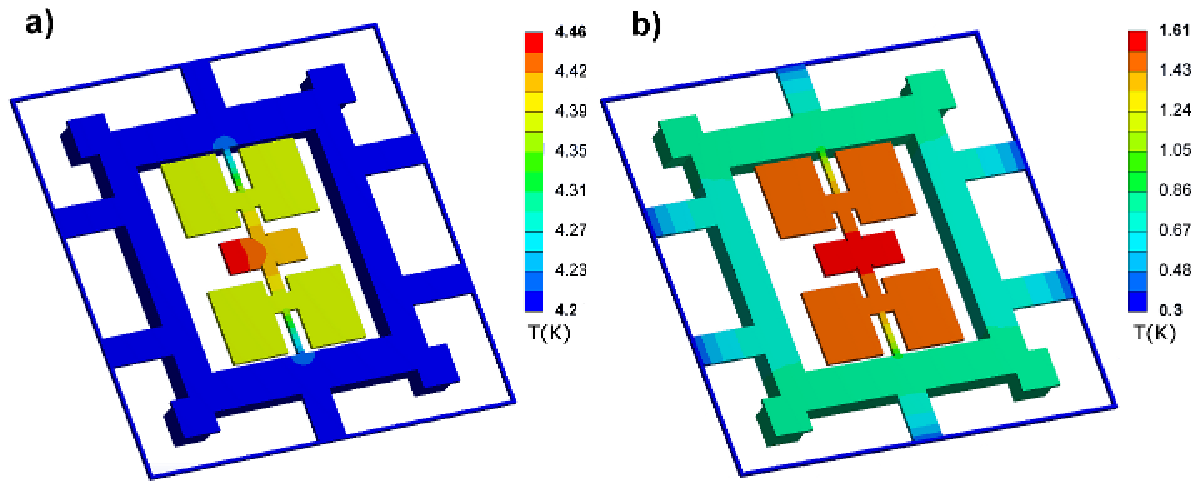}
\caption{\sffamily{Steady-state thermal  analysis  of the QPO}.
FEM simulations showing the effect of the laser beam power absorption. a) Temperature mapping with the background at liquid helium temperature, when the absorbed power is 1 mW. b) Temperature mapping with the background at 300 mK and an absorbed power of 0.1 mW. In both cases the total laser power is applied on the left of the QPO head, on a circular surface of diameter 0.1 mm at the center of the mirror.\label{Temp}}

\end{figure}

An important characteristic that must be evaluated in micro-oscillators is their capability to dissipate the heat produced by the absorbed laser power. This feature actually determines the maximum field amplitude that can be employed in the experiment. In the Figure (\ref{Temp}a) we show the temperature distribution in a DPO when the background (the wafer) is kept at liquid helium temperature and 1 mW of laser power is absorbed in the mirror. The absorbed power in a resonant Fabry-Perot cavity is $\,\frac{4\,\mathcal{T}}{\left(\mathcal{T}+\mathcal{L}\right)^2}\,\mathcal{A}\,P_{in}\,$ where $\mathcal{A}$ is the 
mirror absorption coefficient. For high reflectivity coatings, $\mathcal{A}$ can be below $10^{-6}$ \cite{an}; however, here we consider the more conservative value of $\mathcal{A} = 4\times 10^{-6}$ measured in Ref. \cite{farsi} for the same coating as ours, deposited on a silicon substrate. With this figure, in our cavity an absorption of 1 mW implies an input power as large as 12 mW, with 250 W of intra-cavity power.

The oscillators remains at 4.5 K, thanks to its relatively large thickness and to the high thermal conductivity of silicon (taken from Ref. \cite{klitsner}). Moreover, the temperature is very homogeneous within the oscillators (the spread is less than 0.1 K), with a gradient mainly concentrated in the legs. This feature is important to avoid effects of non-equilibrium thermal noise \cite{conti}. Moreover, it is clear that even a larger dissipated power could be managed.

In the Figure (\ref{Temp}b) we also show a simulation performed with an ultra-cryogenic background (namely, at 300 mK) and a lower absorbed power of 0.1 mW. The sample temperature arrives to 1.6 K, with a moderate improvement with respect to the previous configuration. This is due to the strong dependence of the thermal conductivity of silicon on temperature (in the cryogenic range, it increases roughly as $\sim T^3$). We also remark that in this range the simulation must be considered {\it cum grano salis}, since phononic mean free path is here well over $\sim100 \mu$m and thermal conductivity depends on geometric effects.

\clearpage

\begin{center}
\textbf{\large{Section III. Fabrication of the devices}}
\end{center}

The micro-fabrication process uses SOI wafers made of a $70\pm 1\,\mu$m $<$100$>$ device layer wafer and $400\pm 5\,\mu$m thick handle   wafer, originally joined together with a  $1\,\mu$m buried silicon dioxide by using direct bonding techniques. Both wafers (handle and device layers) are high purity Floating Zone phosphorus doped with a resistivity value greater than 1 k$\Omega\,$cm. The surface roughness RMS (ISO 4287/1) of the device layer is about  0.5 nm, measured  by atomic force microscopy  over an area of  10$\,\mu$m x 10$\,\mu$m. 

The  symmetry axis of the devices are aligned along the $<110>$ crystallographic direction of the device layer wafer. To determine the axis $<110>$, we pattern a wagon-wheel alignment mask in a 200 $\rm{n}m$ thermally grown oxide. This step uses a standard projection lithography technique followed by dry etching of the oxide layer. Wafers were then etched in TMAH to a depth of about 8 $\,\mu$m to expose the crystallographic direction $<110>$. The thermally grown oxide is completely removed from the device layer by wet etching in a BHF/BOE 7:1 solution for  2 min.  

At this stage the back side of the SOI wafer  is spin-coated with a AZ4562 resist with a thickness 10 $\rm{\mu}m$. The mask designed to define the frame structure  is optically aligned with the best-aligned marker in the front side. The full thickness of the handle wafer is then removed by an ICP (inductive coupled plasma) Alcatel DRIE AMS 200 machine based on the Bosch process. The average etching rate is 12.5 $\rm{\mu}m$/min with a temperature of -4 $^\circ \rm{C}$ of the chuck. 

A  fundamental step in the fabrication of the devices is the realization of the highly reflective mirror. To integrate the optical coating deposition with the micro-fabrication process we set-up a lift-off procedure. The device layer is spin coated with a negative high-thick nLOF2070 MicroChemicals photoresist. We obtain a 7.8 $\rm{\mu}m$ thick resist (higher that the mirror thickness) with a spin speed of 3000 rpm. A number of circular regions, corresponding to the position of the mirrors on the device layer, are patterned in the resist, which is then  stabilized  by a soft baking at 120 $^\circ \rm{C}$.   The mirror is obtained by  a deposition of 38 alternate Ta$_2$O$_5$/SiO$_2$ quarter-wave layers (deposited by Ion Beam Sputtering at ATFilms) for a calculated residual transmission of less than 5$\times 10^{-6}$ and a total thickness of about 6 $\mu$m. After the deposition the photoresist sacrificial layer was removed by hot acetone at 90 $^\circ \rm{C}$ and the mirror coating was stabilized with a heat treatment at 430 $^\circ \rm{C}$. RCA cleaning steps are then used to remove residual organic contaminants.

To complete the fabrication of the devices, structures on the device layer are defined with a lithographic step. The front surface is spin coated  with a  layer of AZ4562 resist 10$\,\mu$m thick.  This resist covers all the surfaces and works as protection layer for the mirror areas during the deep-RIE Bosch process. The mask designed to define the front structure is optically aligned with  markers on the back side, with a maximum alignment error of 4 $\mu$m.  
An auxiliary support wafer is fixed on the back side with four photoresist drops hardened by a soft baking. This structure protects the chuck by the contact with the plasma eventually passing through the holes in the structure at the end of the etching process. The full thickness of the device wafer (70 $\mu$m) is removed by DRIE  with an average etching rate is of  7.7 $\rm{\mu}m$/min.
The resist was then stripped using a piranha etch solution and the exposed buried oxide is removed using a BHF wet etch for 15 minutes. In this last step the outer layer of Ta$_2$O$_5$ of the mirror protects the underlying layers avoiding any deterioration of the whole optical coating.

\clearpage

\begin{figure}[bt!]
\includegraphics[width=10cm]{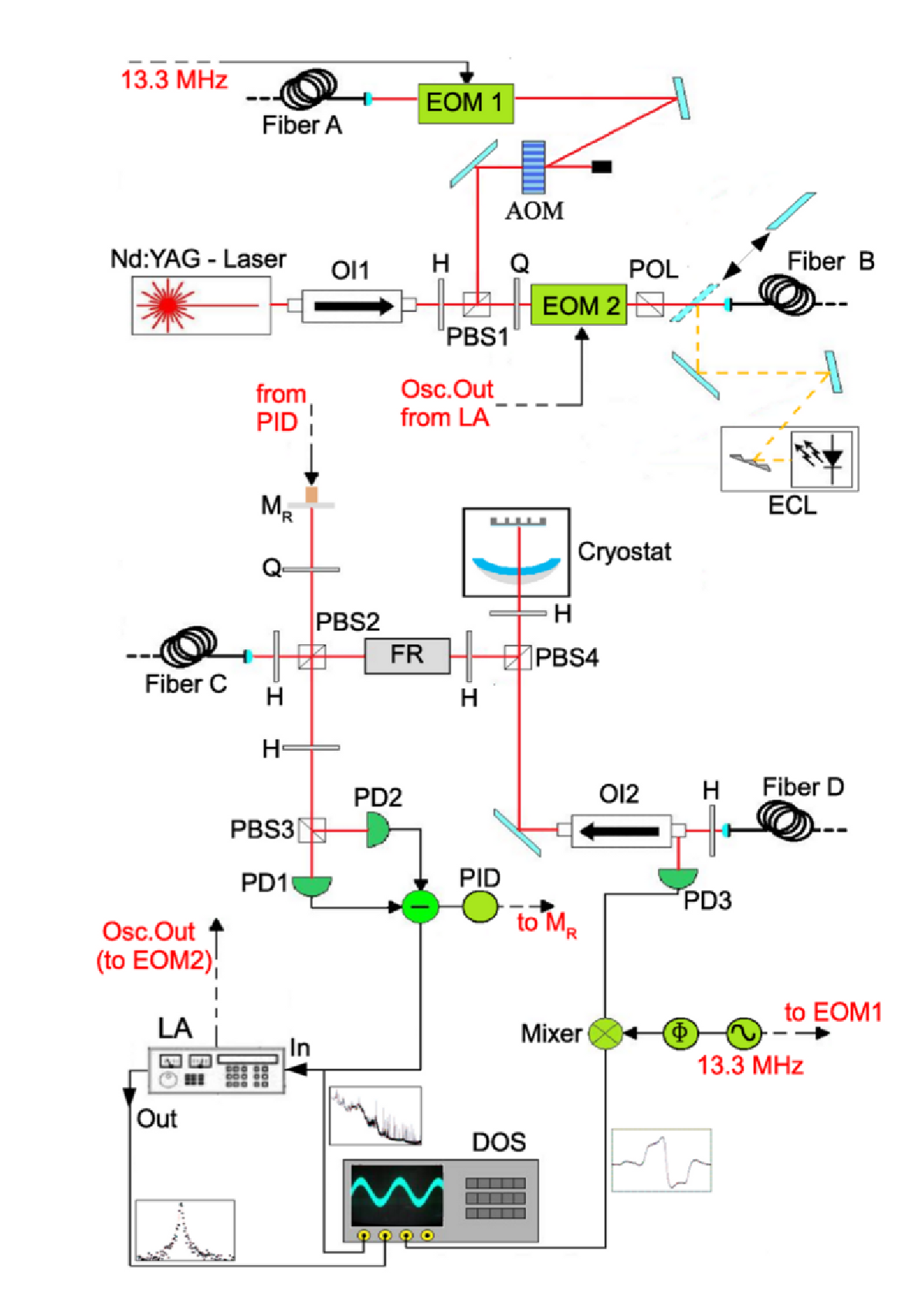}
\caption{\sffamily{Scheme of the experimental apparatus}.
Optical isolator (OI); acousto-optic
modulator (AOM); electro-optic modulator (EOM); half-wave
plate (H); quarter-wave plate (Q);  polarizing
beam splitter (PBS); polarizer (POL); extended-cavity diode laser (ECL); electromagnetically driven mirror ($M_R$); Faraday rotator (FR); photodiode (PD); servo-loop electronics (PID); lock-in amplifier (LA); digital oscilloscope and acquisition system (DOS); delay line for phase control ($\Phi$). Black lines indicate the electronic part of the setup.}
\label{setup}
\end{figure}

%\section{Apparatus}
\begin{center}
\textbf{\large{Section IV. Experimental apparatus}}
\end{center}
The experimental setup is sketched in the Figure~(\ref{setup}). The light source is a cw Nd:YAG laser operating at $\lambda_L = 1064$~nm. After a 40~dB optical isolator, the laser radiation is split into two beams. On the first one (beam A), a resonant electro-optic modulator (EOM1) provides phase modulation at 13.3~MHz with a depth of about 1~rad used for the Pound-Drever-Hall \cite{drever} (PDH) detection scheme. Beam A is then frequency shifted by means of an acousto-optic modulator (AOM), by about 110 MHz. Such shift corresponds to the difference between the resonance frequencies of orthogonally polarized fields, originated by stress-induced birefringence in the Fabry-Perot cavity that is described later. 
The intensity of the second beam (beam B) is controlled by a second electro-optic modulator (EOM2) followed by a polarizer. Both beams are sent to the second part of the apparatus by means of single-mode, polarization maintaining optical fibers, whose terminations can be exchanged. 

After one fiber, the exit (beam C) is aligned in
a Michelson
interferometer followed by a balanced homodyne detection.
In details, a polarizing beam-splitter (PBS2) divides the beam into two
parts, orthogonally polarized, forming the Michelson interferometer arms. At the end of the first one (reference arm) an electromagnetically-driven mirror $M_R$ is used for
phase-locking the interferometer in the condition of maximum displacement sensitivity.
A double pass through a quarter-wave plate rotates by $90^o$ the polarization of the this beam, which is then reflected by PBS2. The polarization of the second arm, sent to the micro-mirror (sensing arm), is instead rotated by
a double pass through a Faraday rotator. The sensing beam is focused with a waist of 80~$\mu$m on the coated oscillator (or mode-matched to the cavity when it is present), and after reflection it is totally
transmitted by PBS2, where it overlaps with the reference beam reflected by $M_R$.
The overlapped beams are then monitored by an homodyne detection, consisting of a half-wave plate, rotating the polarizations by $45^{\circ}$, and a polarizing beam-splitter (PBS3) that divides the radiation
into two equal parts sent to the photodiodes PD1 and PD2, whose outputs are subtracted. The signal obtained is a null-average, sinusoidal function of the path difference in the interferometer. Such a scheme (polarization Michelson interferometer: PMI) is barely sensitive to laser power fluctuations.
The difference signal is used as error signal in the PMI locking servo-loop (the locking bandwidth is about 1~kHz) and also sent to the acquisition and measurement instruments.

The beam exiting from the second fiber (beam D), after an optical isolator, is mode-matched and overlapped to the sensing beam, with orthogonal polarization, in a further polarizing beam-splitter (BS4). The reflected beam is then diverted by the input polarizer of the optical isolator and collected by a fast photodiode for the PDH signal detection. 
The frequency shift between beams C and D, obtained thanks to the AOM,
allows to eliminate any spurious interference and reduce the cross-talk between the two beams in the photo-detection. 

The samples are mounted on a continuous flow $^4He$ cryostat, evacuated down to $10^{-3}$~Pa. The sample holder includes translation stages for building a compact Fabry-Perot cavity. The cryostat thermal shield has two 25~mm diameter access holes that allow to explore several oscillators on the same wafer, when working with the Michelson interferometer. Such holes limit the achievable low temperature to about 10~K, while we have verified that limiting the apertures to few mm$^2$ (thus selecting one single sample) allows to descend below 4.5~K.

After the spectral measurements performed on different oscillators with the Michelson interferometer, we have formed a 0.6~mm long Fabry-Perot cavity using one of the 400 $\mu$m mirrors on the head of the QPO as end mirror, and a $50$~mm radius silica input mirror (its intensity transmittance is $5\times 10^{-5}$). The cavity has a waist of $43 \mu$m, allowing negligible diffraction losses even on the micro-mirror. The same waist could be obtained, with the same input mirror, also with a not too critical near-concentric cavity ($\sim 49.4$~mm long), thus obtaining a cavity linewidth smaller than the mechanical oscillation frequency. The input coupler is fixed on a piezo-electric transducer for cavity tuning. 

For the accurate measurement of the cavity length, we have used an auxiliary extended-cavity semiconductor laser (with optical feedback from a grating in the Littrow configuration), working around 1064 nm. This laser can be course tuned by rotating the grating in a range covering several longitudinal modes of the 0.6 mm cavity, then fine tuned using the supply current and a piezo-electric transducer which translates the grating.

\clearpage

%\section{Measurement techniques}
\begin{center}
\textbf{\large{Section V. Measurement techniques}}
\end{center} 

We have used four different configurations of the experimental setup for measuring the oscillator characteristics. In the first one, for the measurement of the width and area of the mechanical resonance peaks, the fiber input A is connected to the output C (going through the PMI), with the modulator EOM1 switched off and the beam B stopped. The output of the homodyne detection is used to lock the PMI in the position of maximum sensitivity (corresponding to a null average signal) and it is also sent to a digital scope or to a lock-in amplifier followed by the scope. In the second configuration, for the measurement of the oscillator decay time at cryogenic temperature, fiber B is connected to the output D and this second beam (amplitude modulated by EOM2) is used to excite the oscillator by means of radiation pressure. 
In the third configuration, with the micro-mirror used as end mirror in a Fabry-Perot cavity, beam A (with EOM1 switched on) exits through the end fiber D and the PDH signal is used for calibration purposes. Beam B is transmitted to the end fiber C and through the PMI, that is slowly swept. This scheme is used for the measurement of the cavity optical quality. In the fourth configuration, for measuring the cavity length, the beam of the ECL is sent to the cavity through the fiber input B and output C, the reference arm of the PMI is stopped, and beam A is sent to output D. 

\begin{figure}[bt!]
\includegraphics[width=15cm]{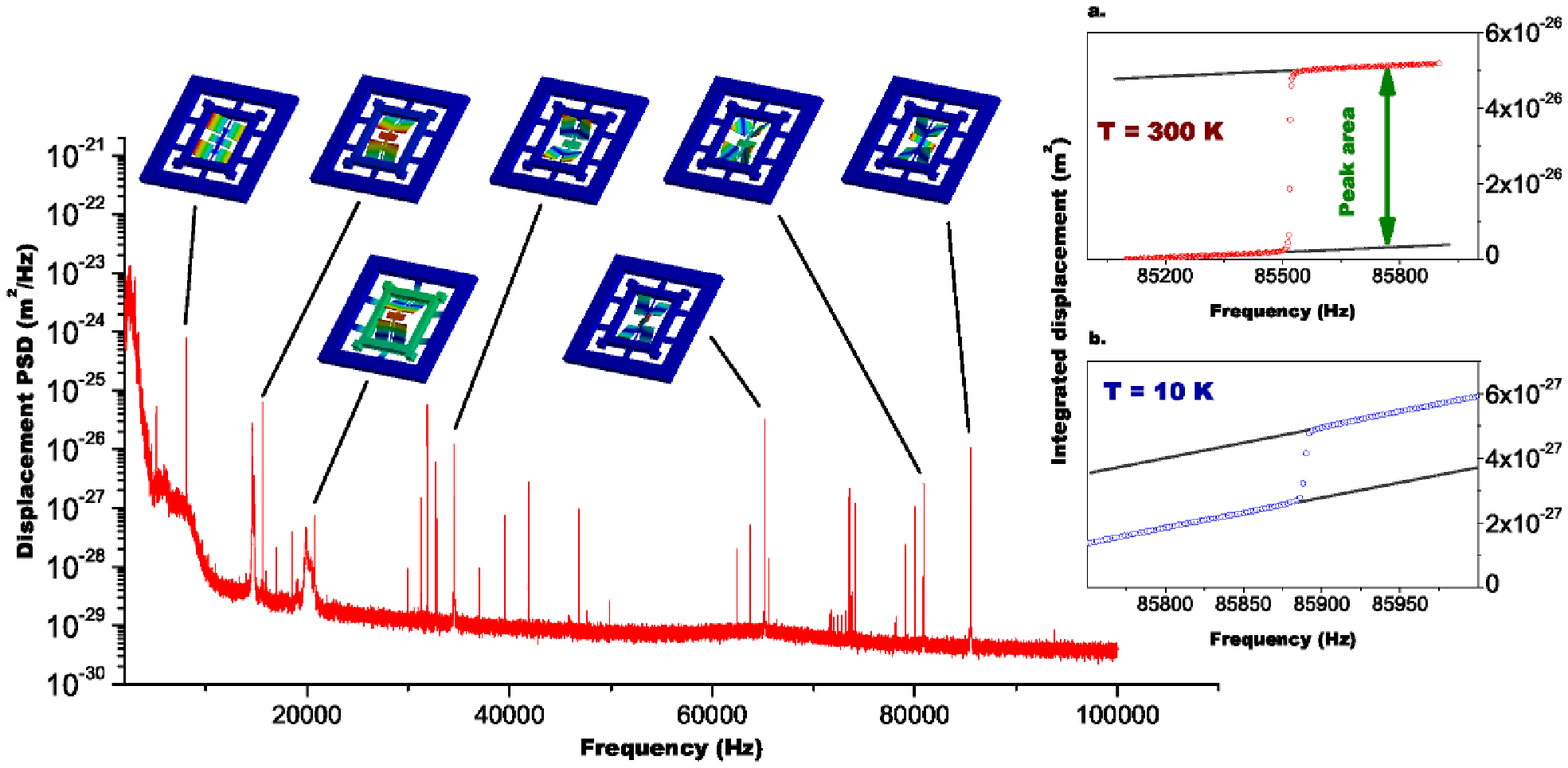}
\caption{\sffamily{Displacement noise spectrum and area below the mechanical peaks.}
The displacement noise spectrum is obtained by phase-locking the Michelson interferometer on 
a dark fringe (condition of maximum displacement sensitivity).
Insets: Integrated displacement around the AS2 mechanical resonance, at room (a) and cryogenic (b) temperature. Couples of dark gray lines correspond to linear fit of the background around resonance, and their distance 
provides the value of the peak area $A$.
\label{figure2_old}}
\end{figure}

Power spectra are calculated and acquired using the integrated Fast Fourier Transform (FFT) analysis software of the digital scope. The measured spectrum $S_V$, in V$^2$/Hz, is calibrated through the
expression $S_{xx} = S_V (\lambda_L/2\pi V_{pk})^2$, where $V_{pk}$ the peak-to-peak value of the PMI interference fringes and $S_{xx}$ is the displacement noise spectrum in m$^2$/Hz. An example of a recorded spectrum is shown in Figure (\ref{figure2_old}). The area of the interesting mechanical peaks is measured by directly integrating over the spectrum, on a region wide several times the peak width, after background subtraction. The procedure is illustrated in the insets of Figure (\ref{figure2_old}). We have verified that the result do not depend on the choice of the FFT windowing and sampling rate. Typical spectra are taken with a record length of 250 kSp and a resolution of 10~Hz (sampling at 2.5 MSp/s).

According to the equipartition theorem, the energy of an oscillator with mass $M$ and displacement $x(t)$ can be expressed as $\,M \Omega_m^2 <x^2>$, where a measurement of $<x^2>$ is provided by the area $A$ below the peak in the displacement power spectral density (PSD). Therefore, one can use the relation $A = \frac{k_B \,T}{M \Omega_m^2}$ ($k_B \,T$ is the oscillator energy) either to extract the effective noise temperature or, assuming that the oscillator is in thermal equilibrium with the environment at temperature $T$, to deduce its effective mass $M$. Our measurements of $A$ have an accuracy of about 30$\%$ (evaluated from their reproducibility). The results are summarized in Table I of the main text, together with the effective mass extracted from $A$ at room temperature. The ratio between the mass values given by $A$ and by FEM simulations, evaluated for the different oscillators, is on the average $1.25\pm0.30$ (the uncertainty reflects one standard deviation), compatible with the experimental uncertainty. Such agreement shows the self-consistency of our approach and justifies the assumption on the thermal origin of the excitation. 
A stronger argument in favor of such assumption is provided by the scaling of $A$ with temperature. Assuming $T=300$ K at room temperature, the temperature estimated from the peak area in the cryogenic environment (also reported in Table I of the main text)  shows a remarkable agreement with the bath temperature of 10~K, measured by two sensors on the cold finger and on the sample holder. Indeed, the temperature extracted from the ratio of the peak areas in the different samples is $10.7\pm2.4$~K, with an uncertainty again compatible with the experimental accuracy.

In order to measure the width of the mechanical resonances, we have used a digital lock-in amplifier, whose internal local oscillator is tuned at 110 Hz from the peak involved in the measurement. The beat note, filtered by the output integrator of the lock-in with a time constant of $640~\mu$s, is then analyzed by the scope with a resolution of 0.1~Hz. An example of the spectrum recorded in this configuration is shown in the left inset of Figure (2) of the main text, together with the fitting function composed of a mechanical resonance plus a flat background. 

For measuring the mechanical quality factor at cryogenic temperature such a resolution is not sufficient. We have therefore adopted a time-resolved detection technique. The excitation beam is amplitude modulated (by means of EOM2) at a frequency very close to the mechanical resonance, for few minutes, then the modulation is switched off and the amplitude of the mechanical oscillation (monitored by the Michelson interferometer) is measured by the lock-in amplifier whose output quadratures are acquired by the scope. The oscillation amplitude is then calculated off-line. An example of the temporal evolution of the oscillation amplitude is shown in the right inset of Figure (2) of the main text, where we also display the fitting exponential decay.

\begin{figure}[b!]
\includegraphics[width=15cm]{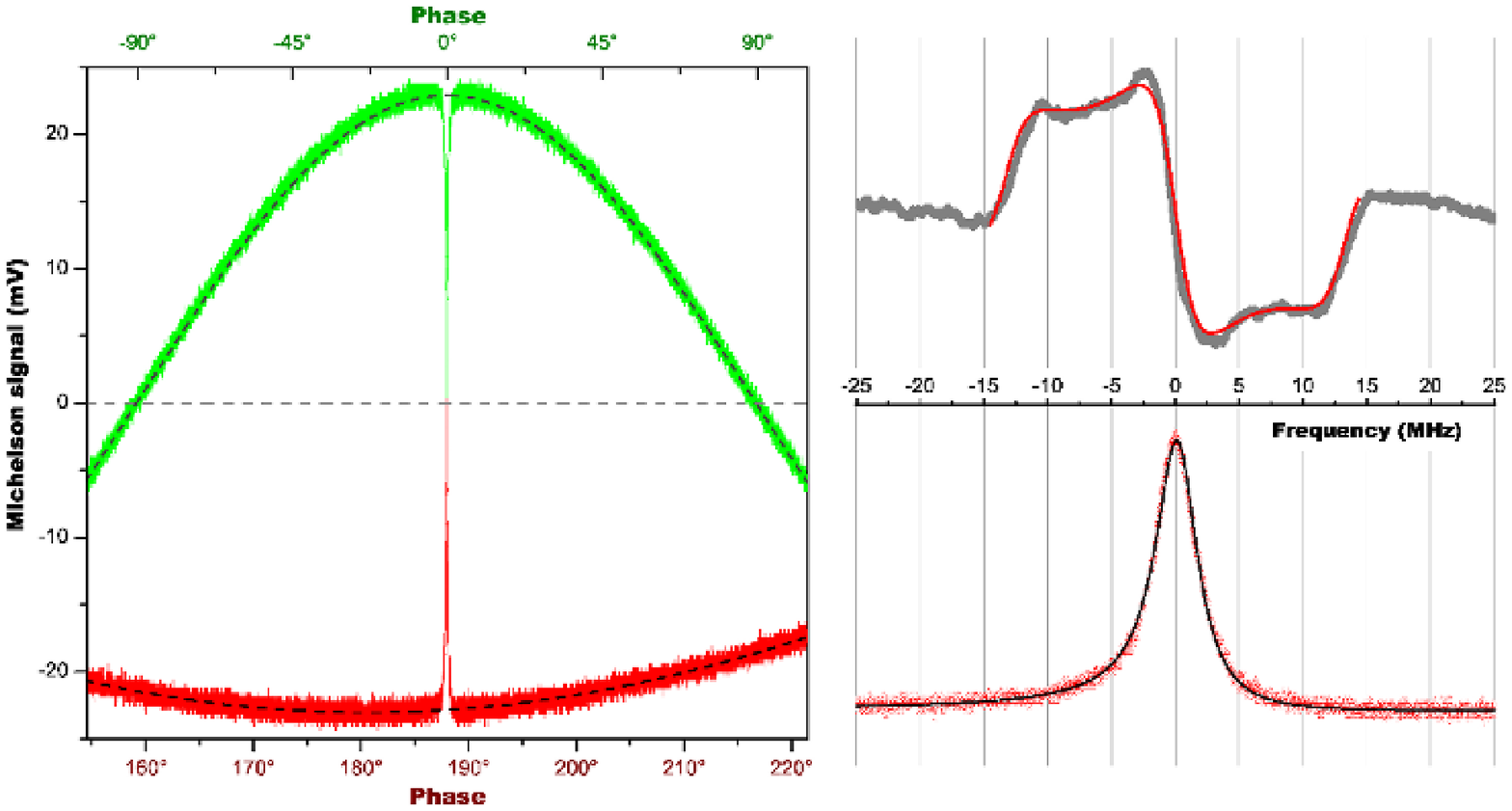}
\caption{\sffamily{Measurement of the Finesse in a Fabry-Perot cavity built with a mirror on the head of a QPO.}   The field reflected from the cavity is measured in a Michelson interferometer, where the cavity resonance appears as a dip on the bright fringe (an upper and a lower bright fringes are shown respectively as green and red traces in the left panel). The depth of the peak gives the coupling coefficient of the cavity, while its width (see on the right an enlarged view of the peak on the lower fringe) gives the cavity linewidth. The frequency calibration in this latter measurement is provided by the distance between the sidebands in a Pound-Drever-Hall  signal (shown above the peak) detected at the same time in a reference laser beam.  \label{figure4_old}}
\end{figure}

The analysis of the cavity optical resonance is performed by scanning its length. In order to avoid any peak shape deformation caused by radiation pressure excitation of the mechanical modes, the laser power is kept as low as possible. Namely, we have used $\sim 3 \mu$W each in beam C and  beam D (measured at the input of the cavity). Instead of recording the dip in the reflected intensity, we have used the PMI to increase the signal-to-noise ratio: the reference beam of the PMI works as local oscillator for the field reflected by the cavity. The amplitude reflection coefficient $\mathcal{R}$ of a Fabry-Perot cavity in the high-Finesse limit can be written as $\,\mathcal{R}=\frac{\mathcal{R}_0 + i \delta}{1+i\delta}\,$ where $\delta$ is the detuning normalized to the half-linewidth of the optical resonance, and $\,\mathcal{R}_0 = \frac{\mathcal{L}-\mathcal{T}}{\mathcal{L}+\mathcal{T}}\,$ is the coupling coefficient. For $\mathcal{R}_0=0$ we have optimal coupling, i.e., the configuration giving the maximum stored energy for given cavity losses and input power (also known as impedence-matched cavity); if $\,-1 < \mathcal{R}_0 < 0$ the cavity is over-coupled. The signal detected
in the PMI is proportional to $\sqrt{I_R I_S} \,\mathrm{Re} \left[(1-\eta+\eta \mathcal{R}_0)\exp(-i \theta)\right]$ where the phase $\theta = \frac{4 \pi}{\lambda_L} \Delta L$ is determined by the unbalance $\Delta L$ between the two arms, with intensities $I_R$ and $I_S$. Normalizing to the fringe amplitude with the cavity out of resonance ($\delta \to \infty$), we can write the PMI signal as
\begin{equation}
S_{PMI} = \cos \theta\,\left[1-\frac{C}{1+\delta^2}\right]\,+\,\sin \theta \, \frac{C\,\delta}{1+\delta^2}
\end{equation}
with $C = \eta \left(1-\mathcal{R}_0\right)$. The coefficient $\eta$ accounts for the mode-matching both in the coupling to the cavity, and between the arms of the PMI. We have estimated both (from the residual power in the cavity transverse modes, and from the fringe contrast seen by the homodyne photodiodes), and they are above 90$\%$.
In Figure (\ref{figure4_old}) we report portions of the PMI signal taken with $\theta$ around $0^{\circ}$ and $180^{\circ}$ when 
$\Delta L$ is slowly swept. The $\theta$ axes are calibrated from fits with a cosine function on signal regions outside cavity resonance (fitting dashed lines are shown in the figure). During the signal acquisition, 
the cavity detuning is scanned around resonance; in the shown portions of the signal the cavity resonance condition $\delta = 0$ happens in near-coincidence with the extremal values of $\cos \theta$, therefore the dip shapes are close to Lorentzian functions. Both dips touch the null value of the signal, visibly demonstrating that the cavity is very close to optimal coupling. The calibration of $\delta$ for each dip is obtained thanks to the PDH signal detected at the same time from beam D (in particular, using the distance between the sidebands). The fits of the dip shapes are then performed using the complete expression of $S_{PMI}$. The fitting procedures give a cavity resonance full linewidth of $3.9\pm0.2$ MHz and a peak depth $C = 0.96\pm0.2$ (uncertainties evaluated from repeated measurements). The linewidth, together with the free-spectral-range of FSR = 252 GHz (see below the description of 
the measurement), corresponds to a Finesse of $65,000\pm3,000$. The expression of $\mathcal{F}=\frac{2\pi}{\mathcal{L}+\mathcal{T}}$,  together with the nominal input coupler transmission of $\mathcal{T}=50\pm5\times10^{-6}$ allows to calculate cavity losses of $\mathcal{L}=47\pm10\times10^{-6}$ and a coupling coefficient $\mathcal{R}_0 = -0.03\pm0.16$. From this last value and the measured $C$ we extract $\eta=0.93\pm0.14$, in agreement with the independent evaluation of the mode-matching.

The cavity FSR = c/2$L_{cav}$ is measured by means of an auxiliary ECL, sent to the cavity together with the Nd:YAG laser beam. The cavity length is slowly scanned over one FSR, and the ECL is tuned in order to be on resonance at the same time as the Nd:YAG (on a different longitudinal mode). The two laser frequencies are then measured by means of a wave-meter with a precision of 100 MHz. From their difference we obtain a FSR of 252 GHz, corresponding to a cavity length of $595 \mu$m. 
 
